\documentclass[acmsmall]{acmart}
\usepackage{subcaption}
\usepackage{float}
\usepackage{cleveref}
\AtBeginDocument{%
  \providecommand\BibTeX{{%
    \normalfont B\kern-0.5em{\scshape i\kern-0.25em b}\kern-0.8em\TeX}}}

\setcopyright{acmcopyright}
\copyrightyear{2018}
\acmYear{2018}
\acmDOI{XXXXXXX.XXXXXXX}

\acmConference[PEARC 2023]{Computing for the common good}{July 23--27,
  2023}{Portland, Oregon}
\acmPrice{15.00}
\acmISBN{978-1-4503-XXXX-X/18/06}

\begin{document}

\title[Enabling knowledge discovery in NHE]{Enabling knowledge discovery in Natural Hazard Engineering datasets on DesignSafe}

\author{Chahak Mehta}
\email{chahak@utexas.edu}

\author{Krishna Kumar}
\email{krishnak@utexas.edu}
\affiliation{%
  \institution{University of Texas at Austin}
  \city{Austin}
  \state{Texas}
  \country{USA}
}


\begin{abstract}
Data-driven discoveries require identifying relevant data relationships from a sea of complex, unstructured, and heterogeneous scientific data. We propose a hybrid methodology that extracts metadata and leverages scientific domain knowledge to synthesize a new dataset from the original to construct knowledge graphs. We demonstrate our approach's effectiveness through a case study on the natural hazard engineering dataset on ``LEAP Liquefaction'' hosted on DesignSafe. Traditional lexical search on DesignSafe is limited in uncovering hidden relationships within the data.
Our knowledge graph enables complex queries and fosters new scientific insights by accurately identifying relevant entities and establishing their relationships within the dataset. This innovative implementation can transform the landscape of data-driven discoveries across various scientific domains.
\end{abstract}





\maketitle

\section{Introduction}
\label{sec:org49148bb}
The pursuit of knowledge discovery is fundamental to the advancement of science. 
In today's research landscape, efficiently managing and analyzing large-scale experimental data has become a pivotal challenge. 
Extracting meaningful insights from such data necessitates not only its efficient storage but also the ability to access and query the data. 
One promising solution to address the complexity of exploring new relations between entities in the datasets is using graph databases, which excel at representing data relationships and dependencies.

We present an innovative implementation \footnote{All code written for the project is open source and can be found at \url{https://github.com/chahak13/tuitus}.
} of extracting parameters and relationships from large datasets by constructing a knowledge graph on heterogeneous scientific data to enable data-driven discovery. 
We build a graph database on the DesignSafe Natural Hazard Engineering Data Depot~\cite{designsafe} at the Texas Advanced Computing Center (TACC). 
Integrating scientific domain knowledge and computing ability is essential for accurately identifying the relevant entities within a dataset to form a knowledge graph. 
This integration is critical for driving scientific discovery. 
One of the main challenges is identifying the precise representation of complex relationships and dependencies underlying scientific phenomena. 
The choice of knowledge representation enables the deciphering of vast amounts of data and transforming it into meaningful information resulting in breakthroughs and deepening our understanding of the world.
Our approach involves a hybrid methodology that not only extracts metadata but also exploits scientific domain knowledge to generate a new dataset that aggregates or synthesizes the original dataset. 
This innovative implementation can advance the frontiers of scientific knowledge discovery by providing a more comprehensive and holistic view of the underlying relationships and dependencies within the data.

We demonstrate the effectiveness of our approach through a case study on the ``LEAP Liquefaction'' datasets~\cite{lbibb22_1, lbibb22_2, lbibb22_3, elghoraiby21_1, elghoraiby21_2, elghoraiby21_3, escoffier23, huang23, ma23, madabhushi23, manandhar23, mitsu23, stone23, takemura23, ueda23, zeghal23} on DesignSafe.
Liquefaction is a natural hazard that involves a complex transition of granular soil from a solid-like response to a fluid-like response with a dramatic loss of strength. 
Discovering the fundamental science that describes this complex multiscale transitionary behavior requires identifying and evaluating potential relationships between different entities in the dataset. 
Although experimental datasets that capture this complex transition exists on DesignSafe, the traditional lexical search on DesignSafe only allows querying only on indexed items, which are often restricted to published metadata, such as the author, title, date, and description.  
The knowledge discovery task with only lexical search is even more challenging when the individual experiments in the datasets, such as ``cyclic direct simple shear,'' does not include the term ``liquefaction'' but is buried within the data when the pore pressure ratio reaches 1.0.
Unless we can identify datasets where that specific condition is reached in the experiment, we cannot create new science. 
Hence, our knowledge graph must support complex queries such as ``Identify cases of loose soil (relative density less than 70\%) where liquefaction was observed (i.e., the pore pressure ratio is 1.0)''.
Enabling such knowledge discoveries requires carefully identifying relevant entities in a dataset and establishing their relationships. 
By creating a knowledge graph on top of large scientific data repositories, we demonstrate the significance of our approach in enabling data-driven discoveries.

\section{Current State of the Art and Challenges}
\label{sec:org14d1606}
\begin{figure}[t]
\centering
\includegraphics[scale=0.25]{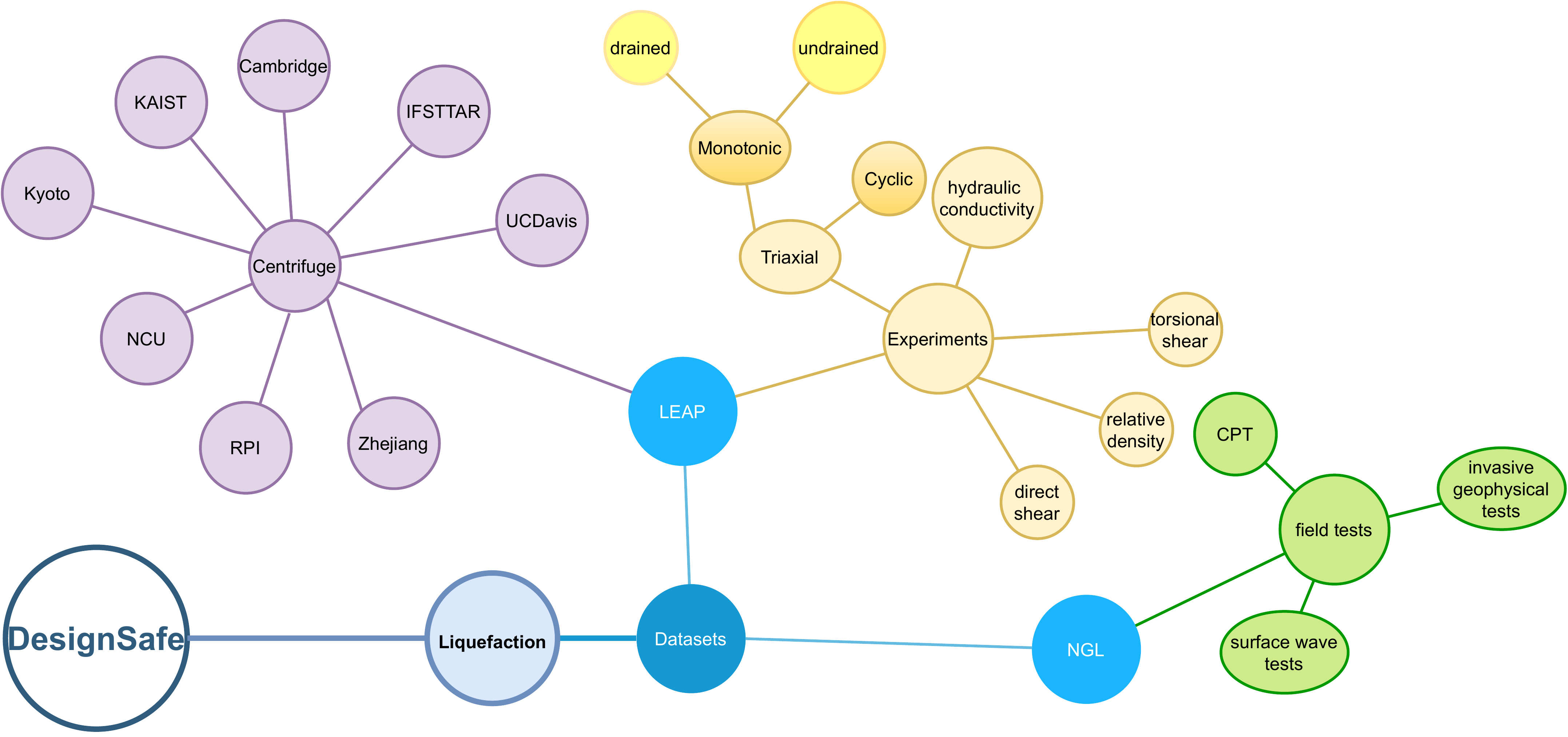}
\caption{\label{fig:knowledge-graph}Knowledge graph of DesignSafe liquefaction datasets.}
\end{figure}

To make data-driven discoveries, it is essential to identify the relevant data that can create new science and models~\cite{buildit}. 
Lexical searchers are limited to indexed terms. Although powerful, text mining and analytics engines like elastic search~\cite{kononenko2014mining} lack semantic understanding as they follow a lexical approach and fail to capture the relationship between entities. 
Graph models are effective in finding relationships or connections between data entities. 
Graph data model inherently captures relationships as first-class citizens, allowing for more powerful and efficient querying of complex connections between data entities.
Resource Description Framework (RDF), a graph model developed by~\cite{BernersLee2001TheSW}, captures the relationship between different entities of the liquefaction dataset on DesignSafe (see~\cref{fig:knowledge-graph}). 
Such a knowledge graph can be constructed using RDF-encoded linked data capable of filtering through and aggregating a diverse set of relevant metadata stored in a graph database~\cite{JWS-facets, kgreview}. 

Identifying and synthesizing data to create new knowledge and relationship requires scientific domain knowledge and a cyberinfrastructure (CI) skillset to extract the relevant information.
Fully automated systems can miss out on crucial domain-specific information, such as failing to create a tag to identify a particular dataset as ``liquefied'' when the pore water pressure ratio column reaches 1.0. 
Summary statistics on the dataset, such as min, mean, and max, will not capture this specific criterion. 
Additionally, the heterogeneous and unstructured nature of the diverse dataset on natural hazards poses another challenge in extracting data from different types and file formats.





\section{Implementation}
\label{sec:org8576338}

We need to identify and ingest pertinent scientific metadata and generate summary statistics to establish data relationships to enable data-driven discoveries. 
Implementing the knowledge graph involves two stages: first, identifying and summarizing the experimental data, and second, creating the graph database using an adapter that processes the generated summaries.

\subsection{Parameters, metadata, and metadata engine}
\label{sec:orgaf7a506}
The first step involves identifying the relevant parameters and metadata for each dataset, defined as a project in the DesignSafe data depot. 
We adopted a hybrid human-assisted approach to defining the metadata tags. 
For this paper, we limited our scope to tabular data in CSV files downloaded from 16 LEAP datasets~\cite{lbibb22_1, lbibb22_2, lbibb22_3, elghoraiby21_1, elghoraiby21_2, elghoraiby21_3, escoffier23, huang23, ma23, madabhushi23, manandhar23, mitsu23, stone23, takemura23, ueda23, zeghal23}, as XLSX files can be more free-flowing and unstructured.
To summarize an experimental result file, we store all the columns of the CSV file as a JSON object along with summary statistics such as the mean, median, count, 25th percentile, and 75th percentile for each numeric column. 
For ordinal columns, we store the frequency of unique values. 
Additionally, we retrieve and store the project metadata to enable standard queries that can be evaluated using relational databases. 
We accomplish this summarization using Python and several libraries, including the DesignSafe \texttt{projects} API, \texttt{agavepy}, \texttt{pandas}, and \texttt{requests}.

\subsection{Graph Database}
\label{sec:org7ee2b01}
The second step involves generating the graph database using the Neo4J graph database platform, deployed on TACC. 
We use Cypher as a graph query language and interact with the database using the \texttt{py2neo} Python library, which acts as an ORM for Python and Neo4J.

\subsection{Knowledge Graphs}
\label{sec:org31f4533}
We define various `nodes' in the database based on the information type and scientific domain knowledge. 
\Cref{table:nodes} lists the various types of nodes and data that we store.
These nodes are then connected by `edges' that show various types of relationships that can be used to create complex queries. 
\Cref{table:relationships} lists the different relationships implemented in the database.

\begin{table}[h]
\caption{\label{table:nodes}Description of Nodes present in the Database.}
\centering
\begin{tabular}{lll}
\toprule
Node & Data in Node & Description\\[0pt]
\midrule
Project & \texttt{project\_id} & Stores the designsafe project ID.\\[0pt]
Author & \texttt{name} & Stores information about PIs.\\[0pt]
Hazard & \texttt{name} & Stores metadata about various hazards.\\[0pt]
Project Type & \texttt{name} & Stores type of project (Experimental, Simulation etc.)\\[0pt]
Experiment & \texttt{uuid}, \texttt{title} & A node that contains experiment metadata.\\[0pt]
Model & \texttt{uuid}, \texttt{title} & Model configuration used in an experiment.\\[0pt]
Event & \texttt{title} & Event information generated for various Models.\\[0pt]
File & \texttt{filepath} & Files generated by Events.\\[0pt]
Data & \texttt{name}, \texttt{mean}, \texttt{median} etc. & A Node for every different type of data stored in the files.\\[0pt]
\bottomrule
\end{tabular}
\end{table}

\begin{table}[htbp]
\caption{\label{table:relationships}Description of Relationships present in the Database.}
\centering
\begin{tabular}{ll}
\toprule
Relationship & Nodes\\[0pt]
\midrule
\texttt{SUPERVISES} & (Author, Project)\\[0pt]
\texttt{NATURAL\_HAZARD} & (Project, Hazard)\\[0pt]
\texttt{TYPE} & (Project, Project Type)\\[0pt]
\texttt{HAS\_EXPERIMENT} & (Project, Experiment)\\[0pt]
\texttt{HAS\_MODEL} & (Experiment, Model)\\[0pt]
\texttt{HAS\_EVENT} & (Experiment, Event), (Model, Event)\\[0pt]
\texttt{HAS\_FILE} & (Event, File)\\[0pt]
\texttt{RECORDS} & (Project, Data), (Event, Data), (File, Data)\\[0pt]
\bottomrule
\end{tabular}
\end{table}

\subsection{Querying the database}
\label{sec:org431f08b}
\begin{figure}
\begin{minipage}[c][5cm][t]{.5\textwidth}
  \vspace*{\fill}
  \centering
  \includegraphics[width=0.5\textwidth]{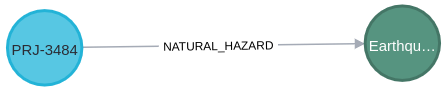}
  \subcaption{Query 1}
  \label{fig:test2}\par\vfill
  \includegraphics[width=0.5\textwidth]{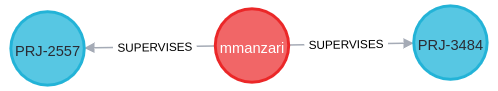}
  \subcaption{Query 2}
  \label{fig:test3}
\end{minipage}%
\begin{minipage}[c][5cm][t]{.5\textwidth}
  \vspace*{\fill}
  \centering
  \includegraphics[width=0.5\textwidth]{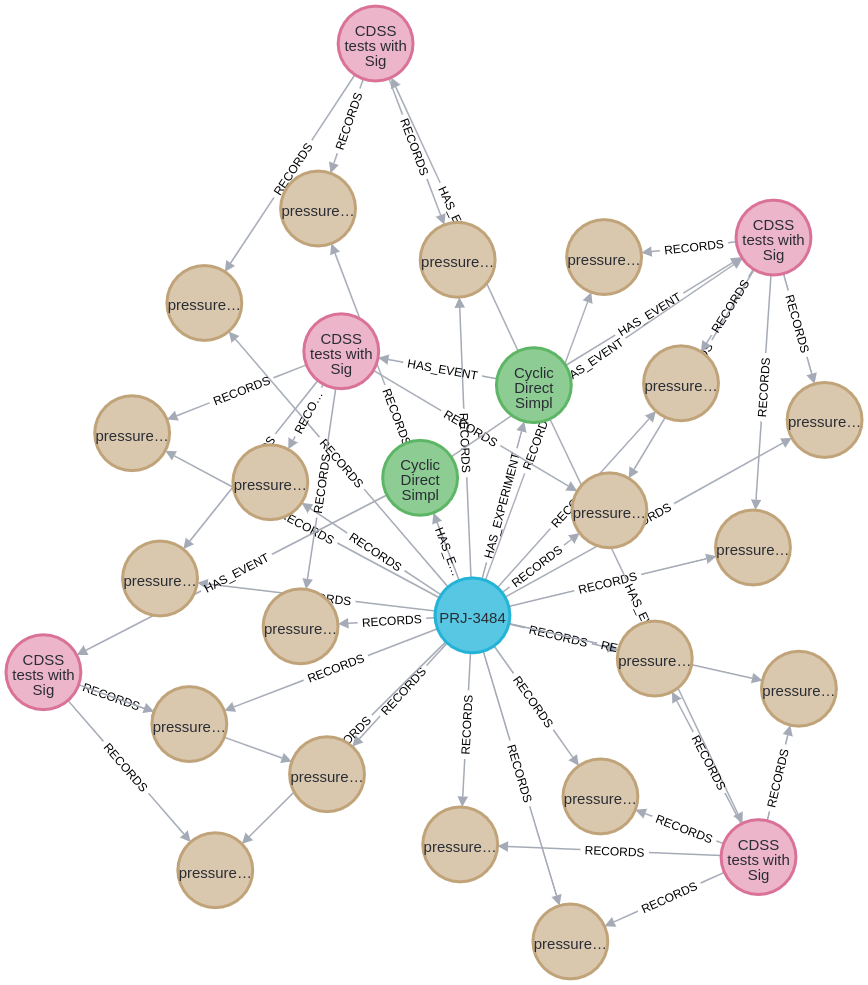}
  \subcaption{Query 3}
  \label{fig:test1}
\end{minipage}
\label{fig:queries}
\caption{Results for Queries described in \cref{sec:org431f08b}}
\end{figure}

We query the database using Neo4J's Cypher Query Language. 
Cypher Query supports simple queries extracting information directly from the metadata (such as project name or title) to more complex queries in identifying behavior under specific conditions. 
Below are a few example queries on the LEAP
Liquefaction database, the results for which are shown in~\cref{fig:queries}.
\begin{enumerate}
\item \emph{Get all projects related to a particular natural hazard type}
\begin{verbatim}
match (p:Project) -[:NATURAL_HAZARD]-> (h:Hazard) return p, h
\end{verbatim}

\item \emph{Get all projects supervised by a specific author}
\begin{verbatim}
match (p:Project) <-[:SUPERVISES]- (a:Author) where a.name = "mmanzari" 
return p, a
\end{verbatim}
\item \emph{Get information about projects with experiments showing imminent liquefaction, i.e., when the results have a mean pore pressure ratio value between 0.5 and 0.7}
\begin{verbatim}
match (p:Project)-->(ex:Experiment)-[:HAS_EVENT]->(ev:Event)-[:RECORDS]->(d:Data) 
where d.name = "pressureratio" and d.mean < 0.7 and d.mean > 0.5 return p, ex,ev,d
\end{verbatim}
\end{enumerate}


\subsection{Extract Metadata using LLMs}
\begin{figure}
    \centering
    \includegraphics[width=0.95\linewidth]{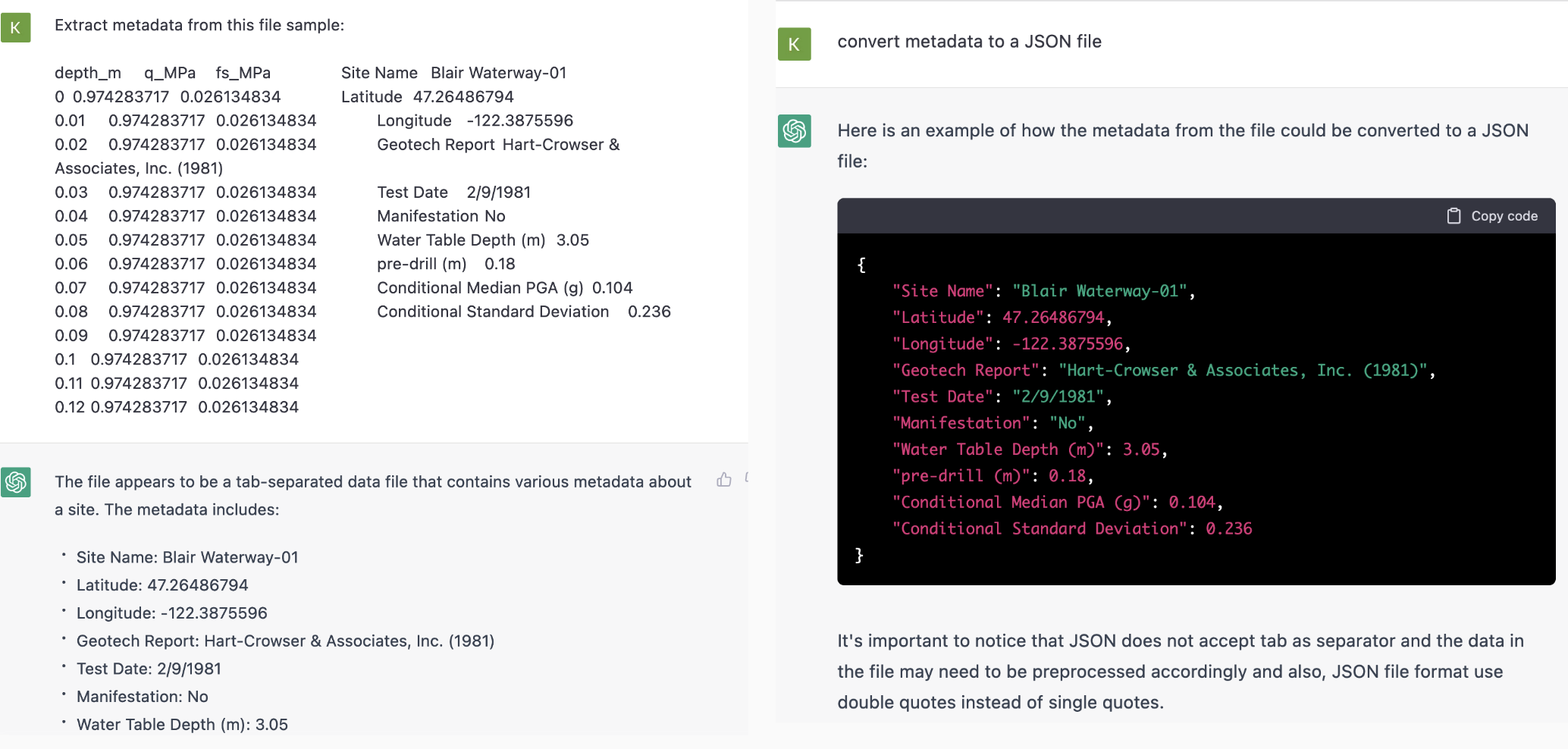}
    \caption{LLM-based data extraction from unstructured XLSX files.}
    \label{fig:llm-xlsx}
\end{figure}

To address the challenge of extracting metadata from unstructured XLSX files, we turn to Large Language Models, such as OpenAI's GPT-4 \cite{openai2023gpt4}. 
Due to the arbitrary structure
of these files, understanding the relevant summaries is a difficult task. 
Our solution involves preprocessing the xlsx file using Python and then using the 
LLM to generate a JSON summary of the file. 
In our initial runs, this approach has shown promising results (see~\cref{fig:llm-xlsx} and~\cite{kumar_krishna_2023_7668209}), and we
are currently working on integrating it into the database engine described earlier. 
As LLMs are probabilistic models, we plan to use a structured prompt to ensure a consistent format for the output that enables us to have the entire pipeline automated.

\section{Conclusion and Future Work}
\label{sec:orgbe5d594}

We developed a hybrid approach by creating a domain-specific data synthesis and metadata extraction model.
By representing the data relationships in datasets of natural hazard engineering, we enable new science and support data-driven discoveries.
The proposed approach is generalizable to other large-scale and heterogenous datasets, enabling new discoveries and knowledge. 

Large-scale data-driven discoveries on complex heterogeneous datasets and natural hazard types would require building extensive relationships across complex experimental, field, and simulation datasets. 
We expect to see knowledge graphs with millions of edges and node clusters. 
DesignSafe currently hosts six petabytes of diverse natural hazard data.
Mining such a large dataset would require close collaboration between domain experts and LLM-assisted metadata extraction. 
Building a knowledge graph on scientific data sets offers the ability to construct data relationships and facilitate knowledge discoveries and push the frontiers of data-driven science. 

\section{Acknowledgments}
\label{sec:orgcd20bd8}
This material is based upon work supported by the National Science Foundation under Grant No. \# 2103937.
Any opinions, findings, and conclusions or recommendations expressed in this material are those of the author(s) and do not necessarily reflect the views of the National Science Foundation. We thank the compute allocations through TACC Frontera and DesignSafe~\cite{designsafe}.

\bibliographystyle{ACM-Reference-Format}
\bibliography{paper}
\end{document}